\documentclass[11pt]{article}
\usepackage{times}
\usepackage{geometry}
\usepackage[utf8]{inputenc}
\usepackage{enumitem,amssymb}
\usepackage{ragged2e}
\newlist{thematic}{itemize}{8}
\setlist[thematic]{label=$\square$}
\usepackage{pifont}
\usepackage{setspace}
\usepackage{eso-pic,graphicx}
\usepackage{tikz}
\usepackage{atbegshi}
\usepackage{url}
\usepackage{xcolor}    
\usepackage{multicol}
\usepackage[english]{babel}
\usepackage{csquotes}
\definecolor{myblue}{RGB}{0,121,194}
\usepackage[
    colorlinks=true,  
    citecolor=myblue,
    linkcolor=myblue,      
    urlcolor=myblue        
]{hyperref}
\usepackage[
    style=numeric,    
    sorting=none,     
    maxbibnames=3,     
    giveninits=true
]{biblatex}
\DeclareNameAlias{default}{family-given}
\DeclareNameAlias{author}{family-given}
\DeclareNameAlias{given}{initials}

\AtEveryBibitem{%
  \clearfield{title}%
  \clearfield{month}%
  \clearfield{day}%
  \clearfield{doi}%
  \clearfield{issn}%
  \clearfield{isbn}%
  \clearfield{note}%
  \clearname{editor}%
}

\AtEveryBibitem{%
}
\renewbibmacro*{in:}{}

\DeclareFieldFormat{eprint}{\href{https://arxiv.org/abs/#1}{\textcolor{myblue}{arXiv}}}

\AtEveryBibitem{%
  \clearfield{month}%
  \clearfield{day}%
  \clearfield{doi}%
  \clearfield{issn}%
  \clearfield{isbn}%
  \clearfield{note}%
  \clearfield{eid}%
  \clearfield{eprintclass}%
}

\DeclareFieldFormat{url}{\href{#1}{\textcolor{blue}{arXiv}}}

\renewbibmacro*{url+urldate}{%
  \iffieldundef{url}{}{%
    \printfield{url}%
  }%
}

\newcommand{\acronym}{FSS}
\newcommand{\kms}{\mbox{$\mbox{\,km\,s}^{-1}$}}

\newenvironment{itemz}
{\begin{list}{$\bullet$}{\setlength{\itemsep}{0pt}}}
{\end{list}}

\begin{document}

\selectlanguage{english}

\raggedright
\Large
ESO Expanding Horizons initiative 2025 \linebreak
Call for White Papers

\vspace{2.cm}
\begin{spacing}{1.6}
\textbf{\fontsize{22pt}{40pt}\selectfont
Mapping dark matter and the emergence of large-scale structure}
\end{spacing}
\normalsize
\vspace{0.5cm}
\textbf{Authors:} Jon Loveday$^1$, Jochen Liske$^2$, Ivan K. Baldry$^3$, Simon P. Driver$^4$, Aaron Robotham$^4$, Sabine Bellstedt$^4$, Luke Davies$^4$, Trystan Lambert$^4$

\vspace{0.3cm}
\textbf{Contacts:} \href{mailto:j.loveday@sussex.ac.uk}{j.loveday@sussex.ac.uk}, 
\href{jochen.liske@uni-hamburg.de}{jochen.liske@uni-hamburg.de},
\href{i.baldry@ljmu.ac.uk}{i.baldry@ljmu.ac.uk}
\linebreak 

\textbf{Affiliations:} \\
$^1$ University of Sussex, Falmer, Brighton, BN1 9QH, UK, \\                 
$^2$ Hamburger Sternwarte, University of Hamburg, Gojenbergsweg 112, 21029 Hamburg, Germany \\
$^3$ Liverpool John Moores University, Liverpool, UK \\
$^{4}$ ICRAR, The University of Western Australia, 35 Stirling Highway, Crawley, WA 6009, Australia \\

\pagenumbering{gobble} 

\pagebreak

\justifying

\section{Background}

The large-scale structure (LSS) of the Universe is shaped by the underlying dark matter (DM) distribution \cite{Davis1985,Frenk1988,Springel2005}.
While dark energy is the dominant contributor to the mass-energy budget today \cite{Aghanim2020},
for the majority of the history of the Universe ($0.5 \lesssim z \lesssim 10^3$) the DM density was dominant, driving the evolution of the Universe, leading to the emergence of LSS, and providing the gravitational substructure in which galaxy formation and evolution, and most of astrophysics, unfolds. 
DM defines the cosmic web --- comprising interlocking filaments, voids, clusters, and groups --- and its distribution encodes critical information about the nature of the dark sector. Different DM models, including cold, warm, fuzzy, and self-interacting variants, predict distinct signatures in the fine-scale structure of the cosmic web, e.g. \cite{Springel2006}.

Theory and simulations are consistent in providing detailed predictions of LSS down to Milky Way masses.
Our capacity to test these predictions empirically is currently lacking, with LSS maps restricted to either very low redshifts (late times), or very small and unrepresentative volumes. Nevertheless, where comparisons can be made the agreement is superb and stands as a testament to the CDM model \cite{Springel2006}.

We now have, or will soon have, well-sampled, wide-area surveys such as the two-degree field galaxy redshift survey (2dFGRS) \cite{Colless2001},
the Sloan Digital Sky Survey (SDSS) \cite{York2000},
the Galaxy And Mass Assembly survey (GAMA) \cite{Driver2011},
the Dark Energy Spectroscopic Instrument (DESI) survey \cite{DESICollaboration2024},
and upcoming Wide Area VISTA Extragalactic Survey (WAVES) \cite{Driver2019}.
Nevertheless, we need to move beyond the local Universe and towards detailed maps at much earlier times. Currently, our studies of LSS are mainly conducted on 4-metre class facilities, which simply lack the capacity to probe to very low mass scales or to significant depth.
The first steps in this direction will shortly come from the innovative ESO MOONs and Subaru PFS facilities, both near-IR capable multiplexed spectrographs on 8m class facilities. They will conduct surveys out to redshifts of 2--3 over areas of a few deg$^2$, and will no doubt transform our understanding of galaxies at these epochs. However, while providing a breakthrough in terms of studying galaxy formation and evolution, they are simply too small in survey volume to map the LSS as a test of CDM.

\begin{figure}[htbp]
\begin{center}
\includegraphics[width=0.49\columnwidth]{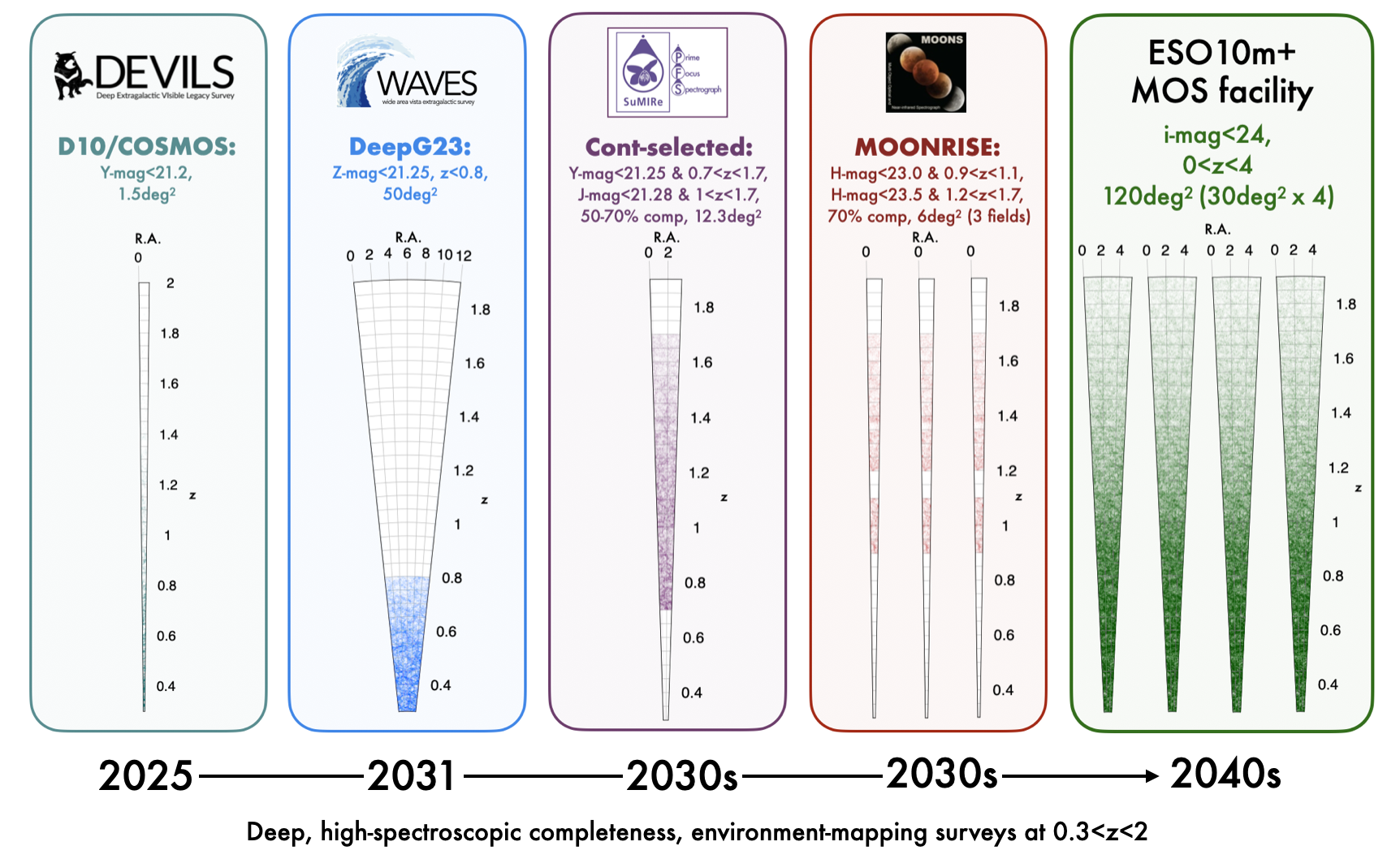}
\hfill
\includegraphics[width=0.49\columnwidth]{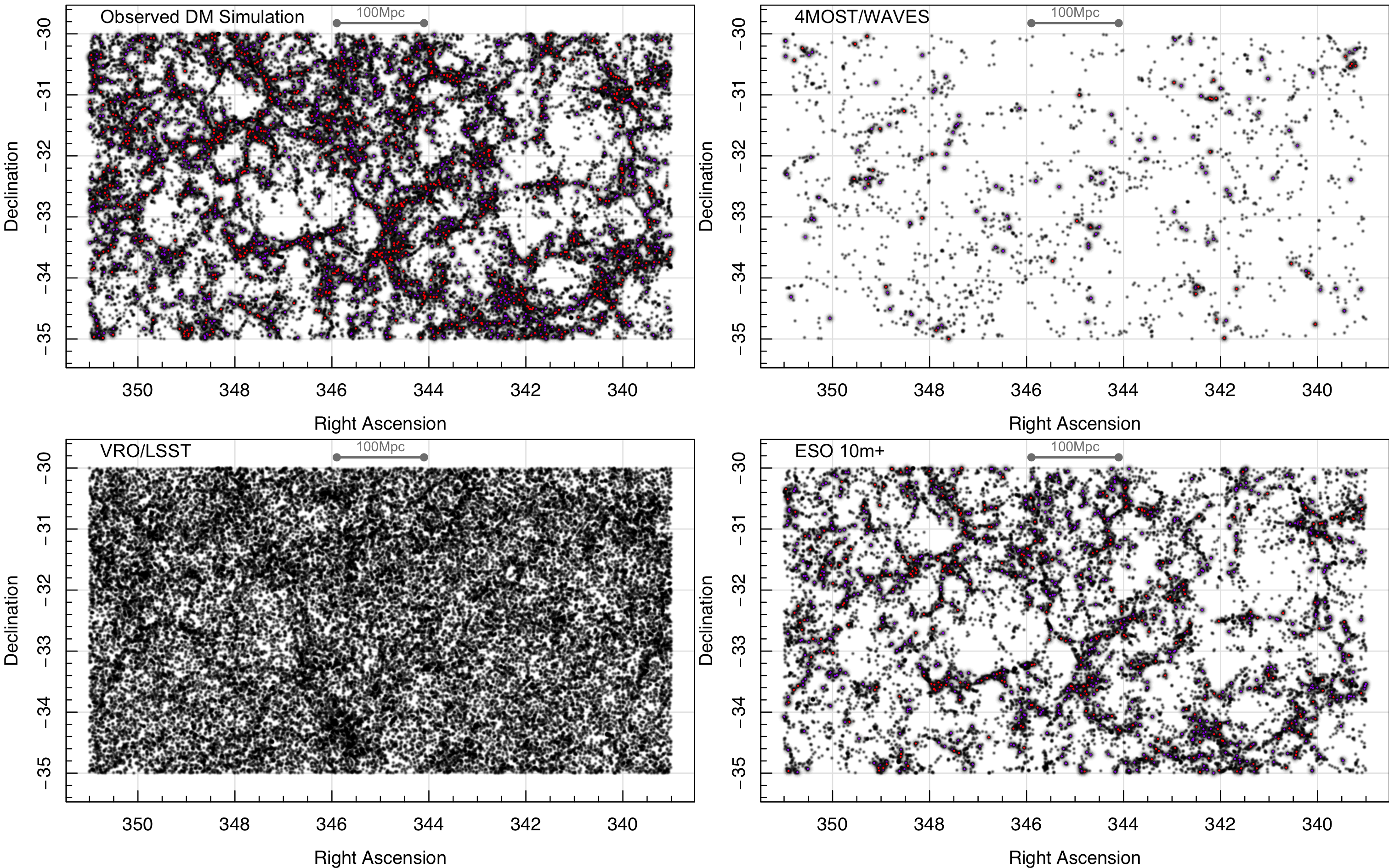}
\caption{Left: Cone plots from existing and planned surveys showing the significant improvement in sampling LSS that can come from a 10m+ wide-area near-infrared multiplexed spectroscopic facility.
Right: Predicted maps of the dark matter distribution at redshift $z=0.2$ as observed by various facilities.  Upper left: the original simulation; lower left: LSST using photo-$z$; upper right: 4MOST WAVES; lower right: \acronym.}
\label{fig:Lightcones}
\end{center}
\end{figure}

Given the critical role of DM in shaping our Universe, it is imperative that we establish the capacity to test the numerical predictions of LSS, with sufficient mass resolution, volume, and depth, to distinguish variations in structure predictions that arise from the nature of DM, such as particle mass, self-interaction, decay, annihilation, and non-negligible baryon/photon interactions.
Critical to measuring the large scale structure and identifying and weighing dark matter haloes are spectroscopic redshifts. While photometric redshifts from broad and narrowband surveys have advanced dramatically, the effective radial distance resolution for defining haloes and mapping large scale structure is still poor (e.g. Fig.~\ref{fig:Lightcones}). For instance, at $z \approx 2$ narrow-band filters can provide a physical resolution of at best 2500 \kms\ whereas typical galaxy groups have velocity dispersions of as little as $\sim50$ \kms\ and the width of a typical filament at around 0.5~Mpc implies a front-back redshift difference of $\Delta z \approx 0.0003$ of 900 \kms. Likewise, to identify galaxy pairs one needs spectroscopic resolution down to 50 \kms\ or $R \sim 6000$.
A 10m+ spectroscopic facility, with wide field of view and high multiplex, with the potential to extend observations to the near-IR, would provide the required capacity.

\section{Approach}

\begin{figure}[htbp]
\begin{center}
\includegraphics[width=0.6\columnwidth]{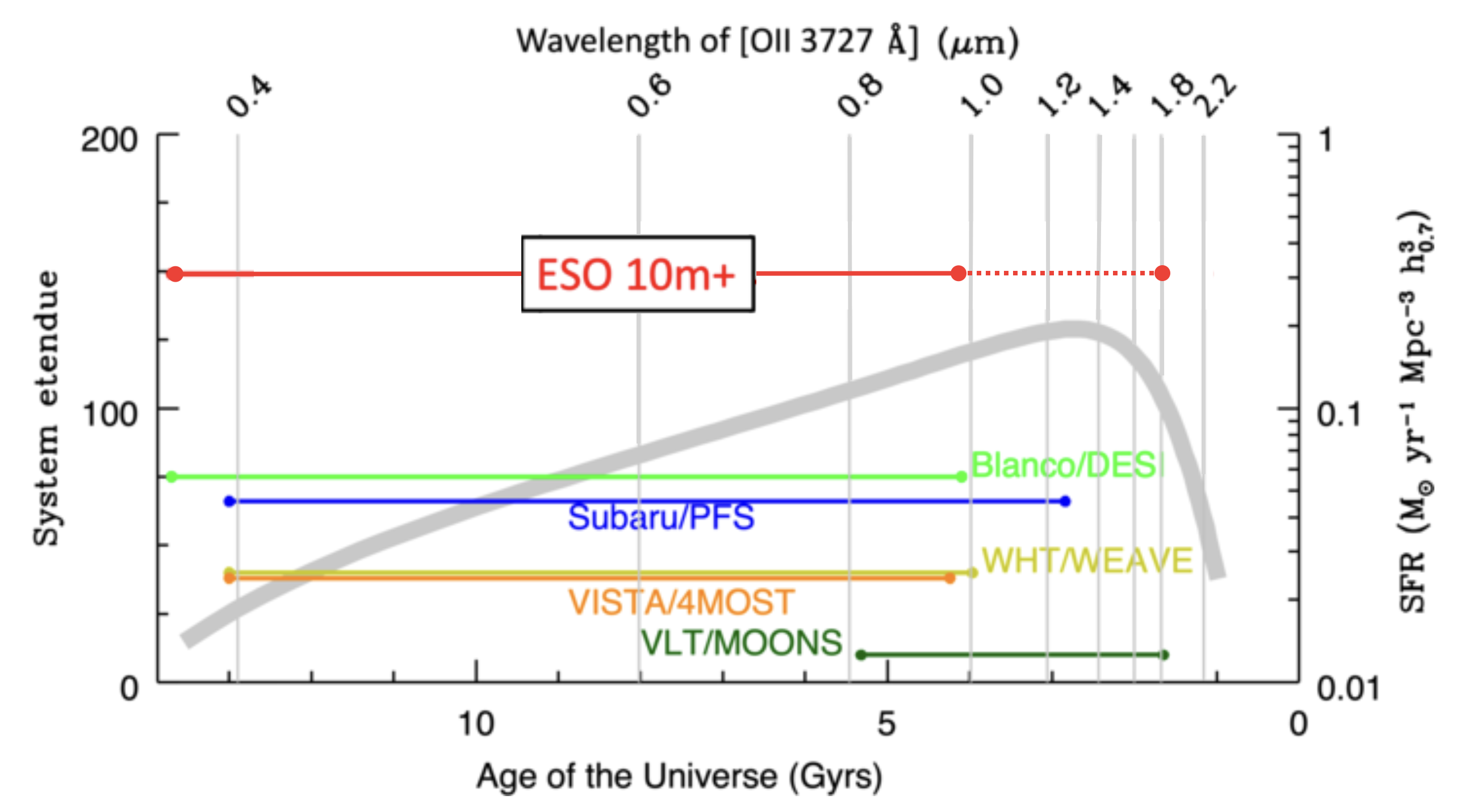}
\vspace{-2mm}
\caption{The cosmic star-formation history (grey band) and the spectroscopic coverage possible from a variety of facilities (including a future $H$-band extension for an ESO 10+m spectroscopic facility). System \'etendue is shown on the y-axis and the vertical lines indicate the observed wavelength of [OII] at different ages.}
\label{fig:grasp}
\end{center}
\end{figure}

Here we outline a future spectroscopic survey, hereafter referred to as \acronym, that would enable an unprecedented exploration of the three-dimensional distribution of DM as traced by galaxy haloes across $\sim$ Gpc$^3$ volumes out to redshift $z \approx 1.5$ in the baseline design and extending to $z \approx 3.5$ with a future spectrograph upgrade to the NIR $H$-band (Fig.~\ref{fig:grasp}).

\acronym\ would observe $\sim 10^7$ galaxies selected to LSST $i < 24$ mag over four 30 deg$^2$ fields to $z \approx 1.5$ over five years, assuming a MOS upper wavelength limit of $0.97 \mu$m, with a later extension to  $z \approx 4.5$ following MOS upgrade to $H$ band.
This would allow us to trace the underlying matter density field that governs the formation and evolution of LSS in the Universe and constitute the most detailed and comprehensive test to date of the $\Lambda$CDM framework and its alternatives, offering a decisive advance in our understanding of the dark Universe.
Highly-complete redshift sampling will allow for the construction of mass maps through group velocity dispersions, enabling model discrimination through measurements of filament thickness, halo abundance, void statistics, and the evolution of the cosmic web.


Furthermore, the relationship between galaxies and their host DM haloes is central to understanding galaxy evolution. Halo mass, assembly history, and environment all influence star formation, morphology, and gas content. \acronym\  will enable the first statistically robust study of galaxy populations as a function of halo mass beyond $z=0.2$, providing critical insights into the co-evolution of galaxies and their dark matter environments.

\section {Requirements}
\subsection{Completeness}

For identifying galaxy groups from which we can derive halo masses via velocity dispersions, every redshift is needed. For example, we would fail to identify the Milky Way (MW) group at high redshift if we missed any of the brightest 3 galaxies (MW, Andromeda, Triangulum), and we would struggle to obtain a credible dynamical mass if we failed to secure 5 redshift measurements (adding LMC and SMC). For the MW group this requires obtaining redshifts to an absolute magnitude of $-17$ mag. A survey operating at less than 90 per cent completeness will essentially miss many MW-like haloes, resulting in a biased sampling of the halo distribution and a biased measurement of the halo mass function. While simulations can be used to make a correction, critical is the ability of simulations to predict fifth brightest member masses accurately. This is currently well beyond the capacity of simulations given the uncertainty around dynamical merger timescales within the halo environment. Hence, the only remedy is to be complete or as close to complete as possible. To probe a volume successfully down to MW halo masses of just a few $10^{12}$M$_{\odot}$, we must survey to high completeness down to $M_i \sim -17$ mag. At $m_i = 24$ mag this equates to a redshift of $z \lesssim 0.35$.

\subsection{Telescope aperture and multiplex}

Current 4-m class facilities, AAT, DESI, 4MOST, struggle to obtain spectroscopic redshifts for  all galaxies at $m_i \approx 21$ within 4 hrs. While some emission line galaxies provide a reliable redshift, quiescent and unusual spectral types require significantly longer. The restriction is a combination of spectral lines shifting out of the optical, $(1+z)^4$ surface brightness dimming, and intrinsically low surface brightness for nearby, low mass systems. 
A 12-m aperture allows a factor of 9 gain in light gathering power or reduced exposure time. 
Hence objects that take 4 hrs on a 4-m will take less than 30 minutes on a 12-m telescope.
Increased mutlitplexing, e.g.\ $2\times 10^4$ fibres, c.f. 1600 4MOST LR fibres, represents a potential further gain of more than 10. Hence a a 12-m, $2\times 10^4$ multiplex facility such as proposed for WST, can represent a factor 100 gain in survey speed. This would allow spectroscopic surveys to extend 3 magnitudes fainter (i.e., $m_i \approx 24$) and achieve a comparable redshift yield per night to 4MOST.

\subsection{Field of view}

1 degree equates to $\approx 60$ Mpc transverse distance at $z \approx 1$. Hence to probe to $500 \times 500$ Mpc scales (necessary to sample the BAO scale at $\sim120$ Mpc and where cosmic variance drops below 5\%) requires surveys that extend over $\sim 100$ deg$^2$. MOONS will cover approximately 4 deg$^2$ within its 100-night extragalactic campaign.  While this represents a major advance, it is clearly insufficient to achieve our goals. The ESO VLTs are fundamentally incompatible with wide-field spectroscopy at the scales required and conversion of a VLT is not possible. Hence a dedicated new 10+m facility is required.

\subsection{Near-infrared capability}

At $z \gtrsim 1.6$ the last obvious and common spectral feature, [OII], lies beyond 9700~\AA\ and hence is visible only in the near-IR. 
As cosmic noon for both galaxies and AGN lies around $z \approx 2$, it is clear that a facility capable of probing to near-infrared wavelengths is essential to study LSS at these redshifts. An extension of spectrograph sensitivity to $H$ band would allow [OII] detection to $z \approx 3.5$.
Note that while UV lines do exist, they are poorly understood, do not occur in all spectra and are more often than not very faint, hence any reliance on UV spectral lines will result in highly incomplete surveys --- possibly suitable for cosmology but not for building halo catalogues.

\section{Summary}
The feasibility of \acronym\ relies on access to a 10+m, wide-field, high-multiplex, spectroscopic facility, ideally with coverage up to $H$ band, such as the proposed Widefield Spectroscopic Telescope (WST).  The key science goals are to:

\begin{itemz}
\item Quantify the halo mass function (HMF) and its evolution since $z = 1$ down to $10^{10}M_\odot$.
This will provide critical constraints on structure formation models and the underlying cosmological framework.
\item Characterize the galaxy stellar mass function (GSMF) down to $10^8M _{\odot}$,
enabling robust tests of galaxy formation efficiency and feedback processes in different environments.
\item Construct a high-fidelity catalogue of $\sim 10^5$ dark matter haloes at $z<0.1$, spanning the mass range $10^{10} M_\odot < M_{\rm halo} < 10^{15}M_{\odot}$. Each halo will be characterized by spectroscopically derived velocity dispersions, enabling mass estimates with better than 20\% precision.
\item Generate direct dark matter maps to test the $\Lambda$CDM paradigm and alternative dark matter models, including warm dark matter (WDM) and self-interacting dark matter (SIDM), through topological and statistical analyses.
\item Establish a legacy database for galaxy–halo connection studies. This resource will support long-term investigations into galaxy evolution, environmental effects, and the baryon–dark matter interplay.
\end{itemz}

\printbibheading
\begin{multicols}{2}
\printbibliography[heading=none]
\end{multicols}





\end{document}